\documentclass[twocolumn,twoside,pra,superscriptaddress]{revtex4}
\usepackage{amsmath}
\usepackage{amssymb}
\usepackage{epsfig}

\usepackage{graphicx}
\usepackage{amsthm}
\usepackage{tikz}
\usepackage{wrapfig}

\newcommand{\ket}[1]{| #1 \rangle}
\newcommand{\bra}[1]{\langle #1 |}
\newcommand{\proj}[1]{\ket{#1}\!\bra{#1}}
\newcommand{\braket}[2]{\left\langle #1| #2 \right\rangle}

\newcommand{\norm}[1]{\| #1 \|}
\newcommand{\tr}{\operatorname{tr}}
\newcommand{\av}[1]{\left\langle #1 \right\rangle}
\newcommand{\eff}{\rm eff}
\newcommand{\nn}{\nonumber}

\newcommand{\mM}{\mathcal{M}}

\newcommand{\mH}{\mathcal{H}}
\newcommand{\mG}{\mathcal{G}}

\newcommand{\cM}{\mathcal{M}}

\usepackage{amsthm}
\newtheorem{theorem}{Theorem}

\begin{document}

\title{Quantum equilibration in finite time}

\author{Anthony J. Short}\email{A.J.short@damtp.cam.ac.uk}
\affiliation{DAMTP, Centre for Mathematical Sciences, Wilberforce Road, Cambridge CB3 0WA, UK}
\author{Terence C.\ Farrelly} \email{tcf24@cam.ac.uk}
\affiliation{DAMTP, Centre for Mathematical Sciences, Wilberforce Road, Cambridge CB3 0WA, UK}

\begin{abstract}
It has recently been shown that small quantum subsystems generically equilibrate, in the sense that they spend most of the time close to a fixed equilibrium state. This relies on just two  assumptions: that the state is spread over many different energies, and that the Hamiltonian has non-degenerate energy gaps. Given the same assumptions, it has also been shown that closed systems equilibrate with respect to realistic measurements. We extend these results in two important ways.
First, we prove equilibration over a finite (rather than infinite) time-interval, allowing us to bound the equilibration time. Second, we weaken the non degenerate energy gaps condition,
showing that equilibration occurs provided that no energy gap is hugely degenerate.
\end{abstract}

\maketitle

\section{Introduction}
It is remarkable that, over one hundred years since its introduction,
quantum mechanics still offers new insights into old problems.
Recently, major progress has been made towards demonstrating that
quantum mechanics alone can justify the methods of statistical
physics \cite{Reimann08,Linden09,Short10,Reimann10,Linden10,Tasaki98,Gemmer04,Popescu06,Goldstein06,Gogolin10,Cramer10,Riera11}, something that had previously required additional postulates.
Here, we will examine what quantum mechanics tells us about
the tendency of systems to equilibrate.

We will focus on extending recent results concerning equilibration over infinite time \cite{Reimann08,Linden09,Short10,Reimann10,Linden10}. These results
apply to general systems that meet just two very weak criteria: non-degenerate
energy gaps (which rules out non-interacting subsystems),  and  high effective dimension (which means that the state is spread over many different energies)
\footnote{For a precise definition of these criteria, see the next section.}.
Using these criteria, it was shown in \cite{Linden09}, that small subsystems of a
large quantum system equilibrate, in the sense that they evolve towards some static state and stay close to it for the vast majority of time.
In \cite{Reimann08}, it was shown that  expectation values of observables
on a closed system also equilibrate.

We will extend these results in two important ways. The first is to prove equilibration over a finite time-interval. This prevents the system from deviating
from equilibrium for arbitrarily long periods of time, and sets an upper bound on the equilibration time. The second is to remove the assumption of non-degenerate energy gaps. Instead, we will prove that equilibration occurs as long as no energy gap is hugely degenerate. This still rules out non-interacting subsystems, but is much more likely to be satisfied by real physical Hamiltonians.
Finally, we will discuss systems with a series of temporary equilibrium states, and the importance of considering  realistic measurements when deriving the Gibbs state.

\section{Setup}
We consider a $d$-dimensional quantum system in the state $\rho(t)$,
evolving unitarily via the time independent Hamiltonian $H=\sum_{n} E_n P_n$,
where $E_n$ are distinct energies, and $P_n$ are the projectors onto the
eigenspaces of $H$ with energy $E_n$. We denote the number of distinct
energies by $d_E$  (note that $d_E \leq d$).
 \footnote{If desired, we could define $d_E$ to be the number of
distinct energies on which $\rho(t)$ has support. This gives a slight
improvement in the results, but has the disadvantage of making $d_E$
and $N(\varepsilon)$ depend on the state, rather than the Hamiltonian alone.}

The crucial feature of the Hamiltonian will be the
energy gaps, rather than the energies themselves. The energy gaps are
given by $E_i-E_j$ with $i\neq j$, and so each gap is labelled by an
ordered pair of integers representing different energy levels.  We
will call this set of labels $\mG$, so that
$\mG=\{(i,j) : i,j \in \{1,2, \ldots d_E\}, i\neq j\}$.
 Thus there are $d_E(d_E-1)$ energy gaps, some of which may be degenerate. For $\alpha\in \mG$ with $\alpha =
(i,j)$, we define the gap $G_{\alpha}=E_i-E_j$.  For clarity, we will use
Greek indices solely to denote energy gaps.

We will be particularly interested in the \emph{density of energy gaps}.
We define $N(\varepsilon)$ to be the maximum number of energy gaps
$G_{\alpha}$ in any interval of size $\varepsilon>0$, meaning
\begin{equation} \label{eqN}
N(\varepsilon) = \max_E |\{ \alpha : \alpha \in \mG, G_\alpha \in [E,
E+\varepsilon) \}|.
\end{equation}
The maximum degeneracy of any energy gap is given by $D_G \equiv \lim_{\varepsilon \rightarrow 0^+} N(\varepsilon)$, and the non-degenerate energy gaps condition corresponds to $D_G=1$. However, note that we will not assume this.

In order to prove equilibration, it is important that the system is  large in a relevant sense. The crucial quantity turns out to be the \emph{effective dimension} of the state, given by
\begin{equation}\label{eqdeff}
d_{\eff}=\frac{1}{\displaystyle\sum_{n}\left(\textrm{tr}\left[P_{n}\rho(0)\right]\right)^2}.
\end{equation}
This measures the number of distinct energies that contribute significantly to the state, which indicates the amount of state-space explored over time.
Note that $1 \leq d_{\eff} \leq d_E$, and that $d_{\eff}=N$ when the state is spread equally over $N$ distinct energy levels.
When the effective dimension is large, the probability of finding the
system in any one energy level is small.  As argued in
\cite{Reimann08}, it is practically impossible to prepare a system
with many particles in a state with a low effective dimension.

Finally, we denote the time average of a quantity over the interval $[0,T]$
by $\av{\cdot}_T$. We also define the infinite time-averaged state
$\omega \equiv \av{\rho(t)}_{\infty}$.

\section{Finite time equilibration of expectation values}
With a time-independent Hamiltonian, there is no way to evolve to an unchanging state
if the system did not start in one.  Furthermore, for systems whose
Hamiltonians have discrete spectra there are recurrences that come
arbitrarily close to the initial state \cite{Bocchieri57,Schulman78}.
Thus, thinking of equilibration as
relaxation to to an unchanging state is not appropriate.  Instead, we
should say
that for a system to be in equilibrium it must spend the vast majority of its time in a state close to some fixed state, where `closeness' is determined by how well we can distinguish the two states.

There are two senses of equilibration in which we will be particularly interested. The first is the equilibration of small subsystems described in \cite{Linden09}. The second is the equilibration of closed systems with respect to some reasonable constraints on our measurement capabilities, described in \cite{Reimann08, Short10}. However, it was shown in \cite{Short10} that both these results can be derived from the equilibration of expectation values. In particular, they can be proved from the following result (Theorem 1 in \cite{Short10}, based on a slightly weaker version in \cite{Reimann08}), which holds for any operator $A$ given the assumption of non-degenerate energy gaps.
\begin{equation}
\label{eq:q}
\left<|\textrm{tr}[\rho(t)A]-\textrm{tr}[\omega
A]|^2\right>_{\infty}\leq \frac{\|A\|^2}{d_{\eff}},
\end{equation}
where $\|A\|$ is the operator norm of $A$ \footnote{The operator norm
is equal to the largest singular value of the operator.}. In  real physical situations, we would
typically expect $d_{\eff} \gg 1$. This means that, for most of the time in
$[0,\infty)$, $\textrm{tr}[\rho(t)A]$ will be close to its average $\textrm{tr}[\omega A]$ relative to the overall scale set by $\|A\|$.

We will now extend this result to show that equilibration happens in a finite time, and also drop the assumption of non-degenerate gaps. Note that we take $\hbar=1$ throughout.

\begin{theorem}
Given a quantum system in the state $\rho(t)$ evolving via a time-independent
Hamiltonian with $d_E$ distinct energies, then for any
operator $A$ on the state space,
any energy $\varepsilon >0$ and time $T >0 $
\begin{equation} \label{eq:theorem1}
\langle |\tr[\rho(t)A] - \tr[ \omega A ]|^2  \rangle_T\\
\!\leq\!  \frac{N(\varepsilon)\norm{A}^2}{d_{\eff}}\bigg(1 + \frac{8 \log_2
d_E}{\varepsilon T} \bigg).
\end{equation}
where $\av{\cdot}_T$ denotes the average over the interval $[0,T]$ and $\omega \equiv \av{\rho(t)}_{\infty}$. The quantities $N(\varepsilon)$ and $d_{\eff}$ are defined in (\ref{eqN}) and (\ref{eqdeff}) respectively.
\end{theorem}

\begin{proof}
We first consider a pure state $\ket{\psi(t)}$ and later extend the
result to mixed states by purification.

If the Hamiltonian has degenerate energies, we choose
an eigenbasis of $H$ such that  $\ket{\psi(t)}$ has non-zero overlap
with only one eigenstate $\ket{n}$ for each distinct energy.
The state at time $t$ is then given by
\begin{equation}
\ket{\psi(t)} = \sum_{n} c_n e^{-i E_n t} \ket{n},
\end{equation}
with $c_n \equiv \braket{n}{\psi(0)}$.
It is clear that $\ket{\psi(t)}$ evolves in the subspace
spanned by $\{\ket{n}\}$ as if it were acted on by the non-degenerate
Hamiltonian $H'=\sum_n E_n \proj{n}$. In this case the
equilibrium state is $\omega = \sum_n |c_n|^2 \proj{n}$, and the effective dimension is $d_{\eff} = \frac{1}{\sum_n |c_n|^4} = \frac{1}{\tr (\omega^2)}$.

Denoting the matrix elements of $A$ by $A_{ij}= \bra{i} A \ket{j}$,
we have
\begin{align}
\langle |\tr&[\rho(t)A]  -\tr[\omega A]|^2 \rangle_T  = \Big\langle
\Big|\sum_{i \neq j}  (c_j^*A_{ji} c_i) e^{-i (E_i-E_j)t}
\Big|^2\Big\rangle_T \nonumber \\
 & =  \sum_{\scriptsize \begin{array}{c} i \neq j \\ k\neq l
\end{array}} \!\!\! \! (c_j^* A_{ji} c_i)  (c_l^* A_{lk} c_k)^* \av{ e^{i[(E_k-E_l) - (E_i
- E_j)]t} }_T.\label{eq:1}
\end{align}
It is helpful to rewrite this expression in terms of energy gaps, by
taking $\alpha = (k,l)$ and $\beta = (i,j)$. We also define the vector
\begin{equation}
v_{\beta}=v_{(i,j)} =  c_j^* A_{ji} c_i
\end{equation}
and the Hermitian matrix
 \begin{equation}
 M_{\alpha\beta} = \av{ e^{i(G_{\alpha} - G_{\beta})t} }_T.
\end{equation}
Adopting a similar approach to \cite{Short10}, equation (\ref{eq:1}) becomes
\begin{align}
\av{ |\tr[\rho(t)A] - \tr[\omega A]|^2 }_T &=  \sum_{\alpha, \beta} v_{\alpha}^* M_{\alpha\beta}  v_{\beta} \nonumber \\
 & \leq  \norm{M} \sum_{\alpha}  |v_{\alpha}|^2  \nonumber \\
 & =\norm{M} \sum_{i\neq j}  |c_i|^2 |c_j|^2 |A_{ji}|^2  \nonumber \\
 & \leq \norm{M} \sum_{i, j}  |c_i|^2 |c_j|^2 |A_{ji}|^2  \nonumber \\
 & = \norm{M} \tr( A \omega A^{\dag} \omega ) \ \nonumber \\
 & \leq  \norm{M}\sqrt{\tr(A^{\dag}\!A\, \omega^2) \tr (A A^{\dag} \omega^2)} \nonumber \\
 &\leq \norm{M}\norm{A}^2 \tr(\omega^2) \nonumber \\
&=\frac{\norm{M}\norm{A}^2}{d_{\eff}}.
\end{align}
The last two inequalities follow from the Cauchy-Schwartz inequality for operators with the scalar product $\tr(A^{\dag} B)$, and the fact that for positive operators $P, Q$, $\tr(PQ) \leq \|P\| \tr(Q) $.
Note that if the Hamiltonian has no degenerate energy gaps, and we consider the infinite-time limit $T \rightarrow \infty$, then $M$ becomes the identity matrix and hence $\norm{M}=1$. In this limit we recover the previous result given by (\ref{eq:q}).

We now consider the general case in which $T$ is finite, and we make no assumptions about the energy gaps in the Hamiltonian. As $M$ is Hermitian, standard matrix norm bounds give  \footnote{In
particular, this follows from $\norm{M}^2 \equiv \norm{M}_2^2 \leq
\norm{M}_1 \norm{M}_{\infty}$, where $\norm{\cdot}_p$ is the matrix
norm induced by the $p$-norm on vectors. For Hermitian matrices $
\norm{M}_{\infty} = \norm{M}_{1} =\max_{\beta} \sum_{\alpha} |  M_{\alpha\beta}| $ \cite{Horn85}.}
\begin{equation}  \label{eqn:norm}
\norm{M} \leq \max_{\beta} \sum_{\alpha} |  M_{\alpha\beta}|.
\end{equation}
The matrix elements of M are
\begin{eqnarray}
M_{\alpha\beta} &=& \frac{1}{T} \int_0^T e^{i(G_{\alpha} - G_{\beta})t}  dt \nn \\
&=& \left\{ \begin{array}{cl} 1 & \textrm{when}\  G_{\alpha} = G_{\beta}  \\
\frac{e^{i(G_{\alpha} - G_{\beta})T} - 1}{i  (G_{\alpha} - G_{\beta})
T} & \textrm{otherwise}. \end{array} \right. \label{eqM}
\end{eqnarray}
Since $ M_{\alpha\beta}$ is an average of quantities with absolute value 1, note that it must satisfy  $|
M_{\alpha\beta}| \leq 1$.

We now break the sum in (\ref{eqn:norm}) into intervals of width~$\varepsilon$, centered on some given energy gap $G_\beta$. There are at most
$N(\varepsilon)$ gaps $G_{\alpha}$ satisfying
$(k+\frac{1}{2}) \varepsilon >  G_{\alpha} - G_{\beta} \geq
(k-\frac{1}{2}) \varepsilon$ for each integer $k$.  In the $k=0$ interval, we take $|M_{\alpha
\beta}|\leq 1$.  For non-zero $k$, we note that $|G_{\alpha} - G_{\beta}| \geq (|k|-\frac{1}{2}) \varepsilon$ and hence from (\ref{eqM}),
\begin{equation*}
|M_{\alpha \beta}|\leq \frac{2}{(|k|-\frac{1}{2})\varepsilon T}.
\end{equation*}

There are $d_E(d_E-1)$ terms in the sum $\sum_{\alpha} |
M_{\alpha\beta}|$, and it is maximised by having as many terms with
small values of $|k|$ as possible.
It follows that
\begin{equation}
\sum_{\alpha} |  M_{\alpha\beta}| \leq N(\varepsilon)  +  2
\sum_{k=1}^{\frac{d_E(d_E-1)}{2}}  \frac{2 N(\varepsilon) }{(k -
\frac{1}{2}) \varepsilon T },
\end{equation}
Where the first term comes from the $k=0$ interval, and the second term comes from the intervals with positive and negative $k$.

Using (\ref{eqn:norm}), and the fact that (for $d_E>1$)
\begin{equation}
\sum_{n=1}^{\frac{d_E(d_E-1)}{2}} \frac{1}{(n-\frac{1}{2})} \leq 2
\log_2 d_E,
\end{equation}
which is proven in Appendix A, we find
\begin{equation}
\norm{M} \leq N(\varepsilon)  \left(1 +  \frac{8 \log_2 d_E}{ \varepsilon T }\right),
\end{equation}
and hence
\begin{equation}
\!\!\langle|\tr[\rho(t)A]- \tr[\omega A]|^2\rangle_T \! \leq\!
 \frac{N(\varepsilon)\norm{A}^2}{d_{\eff}} \bigg(1 + \frac{8 \log_2
d_E}{\varepsilon T} \bigg).\label{eq:X}
\end{equation}

This proves the theorem for pure states. We extend the result to mixed states
via purification \cite{Short10}. Given any initial state
$\rho(0)$ on $\mH$, we can define a pure state $\ket{\phi(0)}$
on $\mH \otimes \mH$ such that the reduced state of the first system is
$\rho(0)$. By evolving $\ket{\phi(t)}$ under the joint Hamiltonian
$H \otimes I$, we will recover the correct evolution  $\rho(t)$ of
the first system.  The expectation value of any operator $A$ for
$\rho(t)$ will  be the same as the expectation value of
$A \otimes I$ on the total system, and $\|A\|=\|A\otimes I\|$.
Note that $N(\varepsilon)$ is the same for $H\otimes I$ as for $H$ and
that the effective dimension of the purified system is the same as the
effective dimension of the original system because
$\textrm{tr}[P_E\rho(0)]=\textrm{tr}[P_E\otimes I \proj{\phi(0)}]$.
\end{proof}

\section{Distinguishability}
Even if Theorem 1 shows that the expectation value of a particular observable equilibrates, this does not necessarily mean that it is impossible distinguish $\rho(t)$ from $\omega$ by measuring that observable \cite{Short10}. Furthermore it  does not tell us that the system as a whole has equilibrated.  For example, when
$\rho(t)$ is pure, for any time $t$ there exists some measurement which can distinguish
$\rho(t)$ and $\omega$ with very high probability $(p_{\rm success} \geq 1-\frac{1}{2 d_{\eff}})$ \footnote{We perform a measurement with projectors $\rho(t)$ and $I-\rho(t)$, and guess that the state is $\rho(t)$ when we obtain the first result, and $\omega$ otherwise.}.

To discuss an appropriate measure of distinguishability, it is helpful to describe a measurement not by its expectation value, but by a Positive Operator Valued Measure
(POVM). Each measurement outcome $a$ is associated with a positive operator $M_a$, such that the probability of obtaining that outcome given the state $\rho$ is $\textrm{tr}[\rho M_a]$. For normalisation, we require $\sum_a M_a = I$ (assuming a finite
number of measurement outcomes).

Suppose we are trying to decide whether a system is in
state $\rho$ or state $\sigma$ by performing a measurement (we will assume that both are equally
likely). The best strategy is to guess whichever state has the highest probability of giving the measurement outcome. For example, if
$\textrm{tr}[\rho M_a]>\textrm{tr}[\sigma M_a]$, we should
guess that we have state $\rho$ when we obtain outcome $a$.
With this strategy, our probability of success is
\begin{equation}
p_{\rm success} = \frac{1}{2} \left( 1 + \frac{1}{2}\displaystyle\sum_a \big| \tr[\rho M_a]-\tr[\sigma
M_a] \big| \right).
\end{equation}
This motivates us to define the distinguishability of $\rho$ and
$\sigma$, given the measurement $M$, to be (see \cite{Short10})
\begin{equation} \label{eq:dist}
D_{M}\left(\rho,\sigma\right) =
\frac{1}{2}\displaystyle\sum_{a}|\textrm{tr}\left[M_a
\rho\right]-\textrm{tr}\left[M_a\sigma\right]|.
\end{equation}
We can go a
step further, and define the distinguishability of
$\rho$ and $\sigma$, given the set of measurements $\mathcal{M}$, to
be
\begin{equation}
D_{\mathcal{M}}\left(\rho,\sigma\right) = \max_{M\in
\mathcal{M}}D_{M}\left(\rho,\sigma\right).
\end{equation}
Naturally, if $\rho =\sigma$, the distinguishability is zero.
Furthermore, it is symmetric and obeys the triangle inequality.
But it may not be perfect: there may be states that are
different but for which the distinguishability is zero. This is not
surprising - the set of measurements may not be good enough to
distinguish any two states.

The special case where $\mathcal{M}$
includes all measurements defines the trace distance, denoted by
$D(\rho,\sigma)$.  It follows that
\begin{equation*}
0\leq D_{\mathcal{M}}\left(\rho,\sigma\right)\leq
D\left(\rho,\sigma\right) \leq 1.
\end{equation*}
The trace distance is a good measure of how similar two states are; in
fact, it is a metric on the set of density matrices because it also
satisfies the property that $D(\rho,\sigma)=0$ only if $\rho = \sigma$.
But this is a good
indicator of why the trace distance may not be suitable from the point of
view of statistical physics: we cannot do any measurement we like on
$10^{23}$ particles, so we will usually miss out on the fine details of the
microstate of the system.

\section{Equilibration of  systems and subsystems}

We can use the notion of distinguishability above to give a more precise definition of equilibration. In particular, we note that as the distinguishability is positive, if its average over some time-interval is small, then it must be small for most times. This leads us to the following definition of equilibration.
\begin{center}
\parbox{8.0cm}{
\textbf{Definition:}
\textit{A system equilibrates to the state $\sigma$ over the interval
$[0,T]$ with
respect to the set of measurements
$\mathcal{M}$ if and
only if
$\left<D_{\mathcal{M}}\big(\rho(t),\sigma\big)\right>_{T} \ll 1$.
}}
\end{center}
If $\left<D_{\mathcal{M}}\big(\rho(t),\sigma\big)\right>_{T}$ is
small, it means that we cannot distinguish
$\rho(t)$ from $\sigma$ with the measurements in $\mathcal{M}$ for
most times in $[0,T]$.

In \cite{Short10}, the equilibration of expectation values is used to prove
equilibration of a closed system given reasonable constraints on the set of possible measurements $\cM$. In particular it is shown that
with no degenerate energy gaps,
\begin{equation}
\label{eq:E1}
\left<D_{\mathcal{M}}(\rho(t),\omega)\right>_{\infty}\leq
\frac{\mathcal{S}(\mathcal{M})}{4\sqrt{d_{\eff}}},
\end{equation}
where $\mathcal{S}(\mathcal{M})$ is the total number of outcomes of all the
measurements that we can do.  This is a huge number, but typically it is
insignificant compared to $d_{\eff}^{1/2}$, \cite{Short10}.  Thus, given the non-degenerate gaps condition,
we would expect equilibration  to occur for realistic measurements on large systems over the infinite time interval $[0,\infty)$.

Here we use Theorem 1 to extend this result to finite time-intervals,
and to Hamiltonians that do not satisfy the non-degenerate energy gaps condition.

\begin{theorem}
Consider the quantum state $\rho(t)$ evolving via a Hamiltonian with
$d_E$ distinct energies. For any energy $\varepsilon >0$ and time $T >0 $, the average distinguishability of $\rho(t)$ from $\omega = \av{\rho(t)}_{\infty}$ over the interval $[0,T]$ using measurements in the set $\cM$ is bounded by
\begin{equation}\label{eq:C1}
\left<D_{\mathcal{M}}(\rho(t),\omega)\right>_{T}\leq
\frac{\mathcal{S}(\mathcal{M})}{4\sqrt{d_{\eff}}}\sqrt{N(\varepsilon)\left(1+\frac{8\log_2d_E}{\varepsilon
T}\right)},
\end{equation}
where $\mathcal{S}(\mathcal{M})$ is the total number of measurement outcomes in $\cM$, and $N(\varepsilon)$ and $d_{\eff}$ are defined in (\ref{eqN}) and (\ref{eqdeff}) respectively.
\end{theorem}

This result and Theorem \ref{th:3} below follow from a straightforward substitution of (\ref{eq:theorem1}) for (\ref{eq:q}) in the derivations given in \cite{Short10}. A detailed proof is given in the Appendix B for completeness.

The main theorem in \cite{Linden09} concerning the
equilibration of subsystems is the result that, again with no degenerate
energy gaps,
\begin{equation}
\label{eq:E2}
\left<D\left(\rho_{S}(t),\omega_S\right)\right>_{\infty}
\leq
\frac{1}{2}\sqrt{\frac{d_{S}^2}{d_{\eff}}},
\end{equation}
where $d_S$ is the dimension of the subsystem, $\rho_S(t)$ is the
state of the subsystem at time $t$ and $\omega_S=\langle
\rho_S(t)\rangle_{\infty}$.  This tells us that small enough subsystems
equilibrate with respect to \textit{all} measurements over
$[0,\infty)$.  This is an extremely strong result.  It says that even
if we can do any measurement we want on a subsystem, which can be
arbitrarily large (as long as $d_S^2 \ll d_{\eff}$), its state is
indistinguishable from $\omega_S$ for most times in
$[0,\infty)$.

Here we can also use Theorem 1 to extend this result to finite time-intervals,
and to Hamiltonians that do not satisfy the non-degenerate energy gaps condition.

\begin{theorem}\label{th:3}
Consider a system evolving via a
Hamiltonian with $d_E$ distinct energies. For any energy $\varepsilon >0$ and time $T >0 $,
the trace distance between the subsystem state $\rho_S(t)$ and
$\omega_S = \av{\rho_S(t)}_{\infty}$ averaged over the interval
$[0,T]$ is bounded by
\begin{equation} \label{eq:C2}
\left<D\left(\rho_S(t),\omega_S\right)\right>_{T}
\leq \frac{1}{2}\sqrt{\frac{d_{S}^2 N(\varepsilon)}{
d_{\eff}}\left[1+\frac{8\log_2 d_E}{\varepsilon T}\right]},
\end{equation}
where $d_S$ is the dimension of the subsystem, $N(\varepsilon)$ and $d_{\eff}$ are defined in (\ref{eqN}) and
(\ref{eqdeff}) respectively.
\end{theorem}

See Appendix C for proof.

To understand how these results compare to the previous ones given by (\ref{eq:E1}) and (\ref{eq:E2}) it is helpful to choose $\varepsilon$ equal to the minimum spacing between (non-degenerate) energy gaps. This means setting
\begin{equation}
\varepsilon = \varepsilon_{\min} \equiv \min_{\alpha, \beta} \{ |G_{\alpha} - G_{\beta}| : G_{\alpha} \neq  G_{\beta} \}.
\end{equation}
For this choice of $\varepsilon$, it is easy to see that $N(\varepsilon_{\min})$ is equal to the maximum degeneracy of a single energy gap $D_G$. Considering this in the context of (\ref{eq:C2}), we obtain
\begin{equation}
\left<D\left(\rho_S(t),\omega_S\right)\right>_{T}
\leq \frac{1}{2}\sqrt{\frac{d_{S}^2 D_G}{d_{\eff}}\left[1+\frac{8\log_2
d_E}{\varepsilon_{\min} T}\right]}.
\end{equation}
This choice of $\varepsilon$ gives the strongest bound in the limit as $T \rightarrow \infty$, yielding a version of (\ref{eq:E2}) which applies to any Hamiltonian.
\begin{equation} \label{inftybound}
\left<D\left(\rho_S(t),\omega_S\right)\right>_{\infty}
\leq \frac{1}{2}\sqrt{\frac{d_{S}^2 D_G}{d_{\eff}}}.
\end{equation}
If the non-degenerate gaps condition is satisfied, then $D_G=1$ and we
recover the previous result. However, this bound shows that subsystems
can equilibrate even with large numbers of degenerate energy gaps.
Note that $D_G < d_E$ for all Hamiltonians, with the maximal value of
$D_G=d_E-1$ being obtained when all energy levels are equally spaced
(e.g. a harmonic oscillator with an energy cut-off). However,
almost any non-trivial Hamiltonian will have a much smaller value of $D_G$.

A good motivation for the non-degenerate gaps condition was to rule out systems composed of non-interacting subsystems. Here, such systems are allowed, but we see that appending a non-interacting ancilla to the original system does not improve the bound given by (\ref{inftybound}). In particular, if the ancilla has $k$ distinct energies $d_{\eff}$ can increase  by a factor of at most $k$, but $D_G$ will also increase  by a factor of at least $k$.

Given that we can prove equilibration in the infinite-time limit for a particular choice of $\varepsilon$, then Theorem 2 tells us that this equilibration will also occur over the finite interval $[0,T]$ as long as
\begin{equation} \label{eq:d}
T \gtrsim \frac{\log_2 d_E}{\varepsilon}.
\end{equation}
We can obtain the best bound on the timescale for equilibration by
choosing $\varepsilon > \varepsilon_{\min}$. If the energy levels in
the Hamiltonian are reasonably evenly distributed over a range $\Delta
E$, and $d_{\eff} \sim d_E$, then a choice of $\varepsilon =\eta \Delta
E/d_E$ (with $\eta \ll 1$) should be sufficient to ensure equilibration for a small subsystem.
This gives a timescale of $T = \frac{d_E \log_2 d_E}{\eta \Delta E}$,
which is much shorter than the recurrence time (which typically grows
exponentially with the dimension \cite{Hemmer58}). Furthermore, we can
find examples which we would expect to take at least time linear in
$d_E$ to equilibrate, so the bound probably cannot be improved
significantly without further assumptions. One such example would be a
timer counting down using basis states of a quantum system of dimension
$d$, then exploding a bomb when it reached 0.

\section{Discussion}
We have extended previous results concerning the equilibration of systems and subsystems, to Hamiltonians with degenerate energy gaps (so long as no gap is hugely degenerate) and to finite equilibration times.

The equilibration times obtained are still very large, but this is inevitable given the generality of the approach. We hope that this work can provide a starting point for proving much shorter equilibration times given further assumptions on the Hamiltonian (such as locality and translation invariance), or on the accessible measurements $\mathcal{M}$ (such as only permitting macroscopic measurements).

Throughout, we have focussed on equilibration to the final (infinite-time) equilibrium state $\omega$. However,  in reality, systems often evolve through a series of temporary equilibrium states over time. For example, consider a hot cup of tea.    Eventually it will cool to room temperature, so it seems to be in equilibrium for the
next hour or so.  However, if we wait a lot longer, it will evaporate,
and then there will be a new equilibrium state over a longer period.
If we wait long enough, even the cup will disintegrate.  Defining the
time-averaged state over the interval $[0,T]$ by $\omega_T =
\av{\rho(t)}_T$, the characteristic timescales for these temporary
equilibrations will correspond to those values of $T$ for which
$\av{D_{\mathcal{M}}(\rho(t), \omega_T)}_T \ll 1$. The analogue of
equation (\ref{eq:C1}) in this case would be
\begin{equation}
\left<D_{\mathcal{M}}(\rho(t),\omega_T)\right>_{T}\leq
\frac{\mathcal{S}(\mathcal{M})}{4} \sqrt{ \frac{N(\varepsilon) \| \tilde{M}\|}{d_{\eff}}},
\end{equation}
where  $\tilde{M} = \av{e^{i[G_{\alpha} - G_{\beta}]t}}_T - \av{e^{iG_{\alpha}t}}_T \av{e^{-i G_{\beta}t}}_T$.
Note that when the system has a series of different time-intervals $[0,T_i]$ over which it equilibrates, and the corresponding equilibrium states $\omega_{T_i}$ are distinguishable, then the timescale for each equilibration must be much longer than the previous one. This is because a state cannot be close to $\omega_{T_i}$ for most of $[0,T_i]$ and $\omega_{T_{i+1}}$for most of $[0,T_{i+1}]$ unless $T_{i+1} \gg T_i$. Furthermore, this means that if a system equilibrates over infinite-time, then the timescale for any temporary equilibration must be much less than that given by (\ref{eq:d}). It would be interesting to study these phenomena more closely in future work, and obtain a strong bound on $\| \tilde{M} \|$.

Another interesting question we have not addressed here is the form of
the equilibrium state $\omega$. In \cite{Riera11}, given additional assumptions, it is shown that the state $\omega_S$ of a small subsystem is close to a Gibbs state $\rho_S^{\rm Gibbs} = e^{-\beta H_S}/\tr(e^{-\beta H_S})$, where $H_S$ is the Hamiltonian of the subsystem and $\beta$ is the inverse temperature. However, one key assumption is that the interaction Hamiltonian between the subsystem and its environment is very weak (satisfying $\|H_{\rm int}\| \ll 1/\beta$). As $\|H_{\rm int}\|$ is extensive and $1/\beta$ is intensive, this seems unlikely to hold for subsystems composed of many particles. We now argue that when the system is a lattice of spins, one should not expect it to equilibrate to a Gibbs state in the sense that $D(\omega_S,\rho_S^{\rm Gibbs}) \ll 1$. This is because we would expect each spin on the boundary of $S$ to be slightly different from its state in $\rho_S^{\rm Gibbs}$ due to edge effects caused by $\|H_{\rm int}\|$. By measuring all of these boundary particles we would generally expect to be able detect these subtle edge effects, and hence distinguish  $\omega_S$ from $\rho_S^{\rm Gibbs}$, implying $D(\omega_S,\rho_S^{\rm Gibbs})\simeq 1$.
In order for standard statistical mechanics to be applied in such cases, it is therefore highly relevant that we cannot make any measurement we like on the system, further motivating the study of restricted measurement sets.

\acknowledgements The authors would like to thank Sandu Popescu for helpful discussions. AJS thanks the Royal Society for their support.
TCF thanks the Robert Gardiner Memorial
Scholarship, CQIF, DAMTP, St John's College and ESPRC for their support.


\begin{thebibliography}{99}

\bibitem{Reimann08}
P.~Reimann,
  ``{Foundation of
  Statistical Mechanics under Experimentally Realistic Conditions},'' {\em
  Phys. Rev. Lett.}, vol.~101, p.~190403, 2008.

  \bibitem{Linden09}
N.~Linden, S.~Popescu, A.~J. Short, and A.~Winter,
  ``{Quantum
  mechanical evolution towards thermal equilibrium},'' {\em Phys. Rev. E},
  vol.~79, p.~061103, 2009.

\bibitem{Short10}
A.~J. Short,
  ``{Equilibration of 
  quantum systems and subsystems},'' {\em New Journal of Physics}, vol.~13,
  no.~5, p.~053009, 2011.

\bibitem{Reimann10}
P.~Reimann, ``{Canonical
  thermalization},'' {\em New Journal of Physics}, vol.~12, no.~5, p.~055027,
  2010.

\bibitem{Linden10}
N.~Linden, S.~Popescu, A.~J. Short, and A.~Winter, ``On the speed of
  fluctuations around thermodynamic equilibrium,'' {\em New Journal of
  Physics}, vol.~12, no.~5, p.~055021, 2010.

  \bibitem{Tasaki98}
H.~Tasaki, ``From quantum dynamics to the canonical distribution: General
  picture and a rigorous example,'' {\em Phys. Rev. Lett.}, vol.~80, p.~1373,
  1998.

  \bibitem{Gemmer04}
J.~Gemmer, M.~Michel, and G.~Mahler, {\em Quantum Thermodynamics: Emergence of
  Thermodynamic Behavior within Composite Quantum Systems}.
\newblock Berlin: Springer, 2004.


\bibitem{Popescu06}
S.~Popescu, A.~J. Short, and A.~Winter,
  ``{Entanglement and the Foundations of Statistical Mechanics},'' {\em Nature Physics},
  vol.~2, p.~754, 2006.

  \bibitem{Goldstein06}
S.~Goldstein, J.~L. Lebowitz, R.~Tumulka, and N.~Zangh\`\i,
  ``{Canonical Typicality},'' {\em Phys. Rev. Lett.}, vol.~96, p.~050403, 2006.


\bibitem{Gogolin10}
C.~Gogolin, ``Environment-induced super selection without pointer states,''
  {\em Phys. Rev. E}, vol.~81, p.~051127, 2010.

\bibitem{Cramer10}
M.~Cramer and J.~Eisert,
  ``{A quantum central
  limit theorem for non-equilibrium systems: exact local relaxation of
  correlated states},'' {\em New Journal of Physics}, vol.~12, no.~5,
  p.~055020, 2010.

\bibitem{Riera11}
A.~Riera, C.~Gogolin, and J.~Eisert,
  ``{Thermalization in Nature and on a
  Quantum Computer}.'' arXiv:1102.2389v3, 2011.


\bibitem{Bocchieri57}
P.~Bocchieri and A.~Loinger,
  ``{Quantum Recurrence
  Theorem},'' {\em Phys. Rev.}, vol.~107, p.~337, 1957.

\bibitem{Schulman78}
L.~S. Schulman, ``{Note
  on the quantum recurrence theorem},'' {\em Phys. Rev. A}, vol.~18, p.~2379,
  1978.

\bibitem{Hemmer58}
P.~C. Hemmer, L.~C. Maximon, and H.~Wergeland,
  ``{Recurrence Time of a
  Dynamical System},'' {\em Phys. Rev.}, vol.~111, p.~689, 1958.

\bibitem{Horn85}
R.~A. Horn and C.~R. Johnson, {\em Matrix Analysis}.
\newblock Cambridge: Cambridge University Press, 1985.

\bibitem{Schwinger60}
J.~Schwinger,
  ``{Unitary Operator
  Bases},'' {\em Proc.\ Natl.\ Acad.\ Sci.\ USA.}, vol.~46, p.~570, 1960.

\end{thebibliography}

\appendix

\section{Proof of Equation (13)}
Here we sketch a proof of equation (13). Note that due to the convexity of the function $1/x$,
\begin{equation}
\sum_{n=2}^{N} \frac{1}{(n-\frac{1}{2})} \leq \int_1^N \frac{1}{x}\, {\rm d}x\, = \ln N.
\end{equation}
Hence
\begin{equation}
\sum_{n=1}^{\frac{d_E(d_E-1)}{2}} \frac{1}{(n-\frac{1}{2})} \leq
f(d_E) \equiv 2 + \ln \left(\frac{d_E(d_E-1)}{2}\right).
\end{equation}
For the cases $d_E=2$ and $d_E=3$ it can be checked explicitly that
$f(d_E) \leq 2 \log_2 d_E$. For $d_E\geq 3$ it can  be seen by
comparing derivatives that $f(d_E)$ grows slower than $2 \log_2 d_E$. Hence we obtain
\begin{equation}
\sum_{n=1}^{\frac{d_E(d_E-1)}{2}} \frac{1}{(n-\frac{1}{2})} \leq 2
\log_2 d_E.
\end{equation}
Also note
that for $d_E=1$, $\langle |\tr[\rho(t)A]-\tr[\omega A]|^2\rangle_T= 0$, so that (\ref{eq:X}) also holds in this  case.

\section{Proof of Theorem 2}
This proof is essentially the same as a proof in \cite{Short10}, but uses the improved results of Theorem 1.

It is helpful to define $\tilde{M}_a
= M_a - \frac{1}{2}I$ for any POVM element $M_a$, such that $\|
\tilde{M}_a\| \leq \frac{1}{2}$.
\begin{eqnarray}
\!\!\!\!\av{D_\cM (\rho(t), \omega)}_T\!  &=&\!  \av{ \max_{M(t) \in \cM} D_{M(t)} (\rho(t), \omega)}_T \nonumber\\
&\leq&\!  \sum_{M \in \cM} \av{ D_M (\rho(t), \omega)}_T \nonumber\\
&=&\!\frac{1}{2}  \sum_{M \in \cM}  \sum_a \av{ | \tr(M_a \rho(t) )  - \tr(M_a \omega) | }_T \nonumber\\
&\leq &\!\frac{1}{2} \! \sum_{M \in \cM}\! \sum_a\! \sqrt{ \av{ \big( \tr(M_a \rho(t) )  \!-\! \tr(M_a \omega) \big)^2 } }_T \nonumber\\
&=&\!\frac{1}{2}\!  \sum_{M \in \cM}\! \sum_a\! \sqrt{ \av{ \big( \tr(\tilde{M}_a \rho(t) )  \!-\! \tr(\tilde{M}_a \omega) \big)^2 } }_T \nonumber\\
&\leq &\! \frac{1}{2}\sum_{M \in \cM} \sum_a \sqrt{\frac{ \|
\tilde{M}_a \|^2 N(\varepsilon)}{d_{\eff}}\left[1+\frac{8\log_2d_E}{\varepsilon T}\right]} \nonumber\\
&\leq &\! \frac{{\cal S}(\mM)}{4\sqrt{d_{\eff}}}\sqrt{N(\varepsilon)\left[1+\frac{8\log_2d_E}{\varepsilon T}\right]},
\end{eqnarray}
where ${\cal S}(\mM)$ is the total number of outcomes for all measurements in  $\mM$.

\section{Proof of Theorem 3}
This proof is essentially the same as a proof in \cite{Short10}, but uses the improved results of Theorem 1.

Consider the orthonormal operator basis for the subsystem given by the $d_S^2$  operators
\begin{equation}
F_{dk_0 + k_1} =\frac{1}{\sqrt{d_S}} \sum_{n} e^{\frac{2 \pi i n
k_0}{d_S}} \ket{(n+k_1)\, \textrm{mod}\, d_S} \bra{n},
\end{equation}
 where $k_0,k_1 \in \{0,1,\ldots d_S-1 \}$ \cite{Schwinger60} and the basis states $\ket{n}$ are an arbitrary orthonormal basis on the subsystem. 

 Then writing $(\rho_S(t) - \omega_S) = \sum_{k} c_k(t) F_k$, and noting that the trace-distance can be defined as $D(\rho, \sigma) = \tr | \rho - \sigma |$, we have 
\begin{eqnarray}
\av{D(\rho_S(t), \omega_S)}_T &=& \frac{1}{2} \av{\tr \big|\sum_k c_k(t) F_k \big|}_T  \nonumber \\
&\leq& \frac{1}{2} \av{ \sqrt{ d_S  \tr \big|\sum_k c_k(t) F_k \big|^2}}_T  \nonumber \\
&\leq& \frac{1}{2}  \sqrt{ d_S \sum_{kl}  \av{c_k(t) c^*_l(t) }_T \tr (F_k^{\dag} F_l )}  \nonumber \\
&=& \frac{1}{2} \sqrt{ d_S \sum_k \av{|c_k(t)|^2}_T} \nonumber \\
&=& \frac{1}{2} \sqrt{ d_S \sum_k \av{|\tr((\rho_S(t) - \omega_S) F^{\dag}_k)|^2}_T} \nonumber \\
\end{eqnarray}
In the second line we have used a standard relation between the 1- and
2-norm \cite{Horn85}. Applying Theorem 1 we obtain
\begin{eqnarray}
\av{D(\rho_S(t), \omega_S)}_T &\leq& \frac{1}{2} \sqrt{ d_S \sum_k \frac{\|F^{\dag}_k \|^2
N(\varepsilon)}{ d_{\eff}}\left[1+\frac{8\log_2d_E}{\varepsilon T}\right]} \nonumber \\
&\leq& \frac{1}{2} \sqrt{ \frac{d_S^2 N(\varepsilon)}{d_{\eff}}\left[1+\frac{8\log_2d_E}{\varepsilon T}\right]}.
\end{eqnarray}

\end{document}